\documentclass[
fleqn,usenatbib]{mnras}

\usepackage{newtxtext,newtxmath}

\usepackage{graphicx}   
\usepackage{amsmath}    
\usepackage[export]{adjustbox}

\title[{Wavelet analysis of the long-term activity of V833 Tau}]{{Wavelet analysis of the long-term activity of V833 Tau}}

\author[R. Stepanov et al.]
{R.~Stepanov,$^1$ \thanks{E-mail: rodion@icmm.ru}
N.I. Bondar',$^{2}$
M.M.~Katsova,$^3$
D. Sokoloff\,$^{4,5,6}$ and
P. Frick$^{1}$\\
$^1$ Institute of Continuous Media Mechanics, Korolyov str. 1, 614013 Perm, Russia\\
$^2$Crimean Astrophysical Observatory RAS, Nauchny, 298409 Crimea\\
$^{3}$ Sternberg State Astronomical Institute, Lomonosov Moscow State University, Moscow, 119991, Russia\\
$^{4}$IZMIRAN, Kaluzhskoe shosse, 4, Troitsk, Moscow,  142191, Russia\\
$^{5}$Department of Physics, Moscow State University, Moscow, 119991, Russia\\
$^{6}$Moscow Center of Fundamental and Applied Mathematics, Moscow, 1199991, Russia\\
}

\date{Accepted .... Received ...; in original form ...}

\pubyear{2020}

\begin{document}
\label{firstpage}
\pagerange{\pageref{firstpage}--\pageref{lastpage}}

\maketitle

\begin{abstract}
The bulk of available stellar activity observations is frequently checked for the manifestation
 of signs in comparison with the known characteristic of solar magnetic modulation.
The problem is that stellar activity records are
 usually an order of magnitude shorter than available observations of solar activity variation.
Therefore, the resolved time scales
 of stellar activity are insufficient to decide reliably that a cyclic variation for a particular star
 is similar to the well-known 11-yr sunspot cycles. As a result, recent studies report several stars with
double or multiple cycles which serve to challenge the underlying theoretical understanding.
 This is why a consistent method to separate 'true' cycles from stochastic variations is required.
In this paper, we suggest that a conservative method, based on the best practice of wavelet analysis
 previously applied to the study of solar activity, for studying and interpreting the longest available
stellar activity record -- photometric monitoring of  V833 Tau for more than 100 years.
We find that the observed variations of V833 Tau with timescales of 2 -- 50~yr should be comparable
with the known quasi-periodic solar mid-term variations, whereas the true cycle of V833 Tau, if it exists,
should be of about a century or even longer. We argue that this conclusion does not contradict
the expectations from stellar dynamo theory.
\end{abstract}




\section{Introduction}

Spatial and temporal variations in the activity of stars and, especially,
the Sun are the obvious manifestations of complicated processes in turbulent magnetoconvection.
The understanding of
mechanisms that underlie many of these processes is still
a major open issue in astrophysics.
The dynamo problem remains a key challenge despite the fact that
the pioneering formulation of the principal ''mechanism`` \citep{Parker55} has long
been developed theoretically and verified experimentally. There are also
many related problems of theoretical and practical interests, for example,
 coronal heating (see for review \cite{Testa15}), stellar flare oscillations \citep{2018MNRAS.475.2842D}
 or seismic signatures \citep{2016A&A...596A..31S,2019FrASS...6...52K}.
A comparison of stellar and solar activity is a natural line of research \citep{2008ssma.book.....S}.
The possibilities of observing solar activity are significantly superior in accuracy and in
the variety of available tools than those associated with
measuring stellar activity. The study of a wide population of Sun-like stars
has the advantage of providing a distribution of control parameters \cite{2014MNRAS.441.2361V},
which makes results more meaningful and robust for assessing various theories and models.

Important systematic observations of stellar activity were accomplished by the well-known HK project
 at the Mount Wilson Observatory (MWO) that began in 1966 and lasted over three decades
\citep{Baliunas1995ApJ}. Chromospheric Ca II H+K emission in narrow passbands were measured for
almost 100 stars. Other programs at the Lowell Observatory (1984-2000) and
Fairborn Observatories (1993-2003) continued studying the (year-to-year)
Strömgren b, y photometric variability of a sample of 30+ stars
\citep{1997ApJ...485..789L,1998ApJS..118..239R}.
The Solar-Stellar Spectrograph  is a project (since 1994)
dedicated to long-term observations of the Sun and Sun-like stars with
the same instrument at Lowell Observatory  \citep{1995ApJ...438..404H}.
Recent space-based asteroseismology missions such as MOST \citep{2003PASP..115.1023W},
 CoRoT  \citep{2009A&A...506..411A}, and Kepler \citep{2010Sci...327..977B}, as well as
ground-based networks like the Stellar Observations Network Group (SONG; \cite{2008IAUS..252..465G}),
contribute a lot of complementary data for solar-type stars.

The well recognized  11-yr solar cycle (also known as the Schwabe sunspot cycle)
is a distinctive  timescale in comparison with other time scales of solar activity \citep{hathawayLR}.
It is natural to attempt to identify pronounced cyclic activity in other stars.
However, available observational data for any particular star remained much smaller than that
for the continuous observation and study of sunspot activity.
This observational deficiency indeed
complicates any direct solar-stellar comparison.
The telescopic recordings of sunspot activity
available for the past four centuries tell
us that the
observation of any particular cycle or even pair of cycles
would not be sufficient to develop the comprehensive
dynamo model which we have so far.
Solar dynamo as a physical process representing the underlying solar magnetic activity includes
a dominant eigen-mode of corresponding mean-field equations as an explanation of the 11-yr cycle.
 Various nonlinear effects or purely turbulent variations cause significant amplitude and period modulations
 of the 11-yr cycle \citep{Frick2020}. In principle, spherical dynamos can include several dynamo active shells
and manifest a transition from axi- to nonaxisymmetric dynamo modes \citep{2018A&A...616A.160V}.
However, it is a very demanding assumption and one shell can enslave another one
 and hence reducing the number of independent eigen-modes \citep[e.g.][]{2007MNRAS.377.1597M}.
 Penetration of the activity wave from the lower shells up to the surface is another difficulty associated
with the available
schemes of solar activity \citep[e.g.][]{2011A&A...531A..43M}.
The observation of non solar-like stellar activity is a motivating reason for considering the
dynamo model with anti-solar differential rotation, in which the equator rotates
 slower than poles \citep{2020MNRAS.491.3155K}.
The identification of long-term stellar activity variations with dynamo cycles is of crucial importance for their
interpretation and modelling \citep{2018MNRAS.478.4390L}.

However, the stellar data available for
some stars demonstrate clear quasi-stable cycles (e.g., HD81809 or HD10476 in \cite{Frick2004}),
whereas other data illustrate a periodicity that
 varies substantially even during the full records of
 several oscillations (e.g., HD201092, HD219834A in \cite{Frick2004}).
It is difficult to conclude that a given set of observations identifies a clear solar-type cycle
if only two or three cycles are covered \cite{2017A&A...598A..77K}.
Moreover, a solar-stellar comparison becomes more complicated due to
the necessity to take into account stellar mass, age, rotation rate and orientation of the rotation axes.
Having no solar ``identical twin'', we cannot be sure that the observed stellar activity is
 truly another ``solar experiment''. The Kepler mission has provided a data sample of 4000 Sun-like stars
with measured rotation periods. About 10\% of these stars have long-term variability in
their observed fluxes.  Although the full records did not exceed 4.5 years, nevertheless one can identify
a sample of stars with apparently complete cycles \citep{2017ApJ...851..116M,2019MNRAS.485.5096K}.

One more point here is that the main bulk of the examined stellar cycles has
the length of the order of the Schwabe cycle or shorter.
It looks like a selection effect will be introduced
 because of the limited length of the available observational time series.
 Therefore, it is essential to have a method that may allows one to isolate the cases
where one can deduce at least a lower estimate for a long-term activity cycle period
 despite the fact that the true periods cannot be accessible just yet from
the available observations \cite{2016A&A...590A..11V}.
The desired method should explain why the shorter time scales presented in the time series of interest
should be considered as something else rather than the main activity period comparable with the Schwabe cycle.
It is hard to conclude that a star is exhibiting a type of Maunder Minimum like activity phase
 based on observations of low activity over a short time span  \citep[cf.][]{2009AJ....138..312H}.

We note that even the 400-year period of
instrumental observations of sunspot activity has not finalized the discussion
concerning the physical status of various time scales isolated in the solar activity record,
and that the cycles supplementary to the main Schwabe cycle are still debatable.
 We emphasized here that the example of a star with a record of 10 time scales can be more valuable
 than 10 records of different stars with a length of one time scale. In fact, we have patterns of variation
for the Sun and 72 sun-like stars over 1-3 decades but its solar-like behaviour is
still questionable \citep{2018ApJ...855...75R}.
The aim of this paper is to develop a conservative method for
the interpretation of the multi-timescale stellar activity using wavelet based techniques.
As an instructive example, we analyzed the data of the longest available stellar activity --
photometric monitoring of  V833 Tau. These data can provide a first direct evidence that the star
may have an activity period of about a century long.

\section{V833 Tau: general properties}
V833 Tau (GJ 171.2A or HD 283750) is a single-lined spectroscopic binary; the second component
was detected by \cite{Heintz1981ApJS}.
The orbital elements of the system are determined by \cite{Getal85}, and described
in more detail by \cite{Halbwachs2000IAUS,Halbwachs2018AA}.
The second star (HD 283750b) is moving on the circular orbit with a period of 1.788011 d,
 its mass is estimated to be about 0.17 $M_\odot$ or less, and it is consistent with being
a brown dwarf. The infrared spectroscopic study by \citet{LucasRoche2002} allows one
to consider HD 283750b as a hot Jupiter planet with mass of 50 $M_{\rm Jup}$.
\cite{Getal85} presented the history of the study of HD 283750.

\cite{Eggen1965ApJ} found that the brightness of the star CJ171.2 A is $V=8.42^m$ and
the spectral type is dK5pe with emission in hydrogen lines.
Its photometric and spectroscopic features correspond to the BY Dra type variables \citep{Bopp1981ApJ}.
The rotation modulation with the period of 1.85 d
was found by \cite{Pettersen1989AA}, and later this value was refined by
photometric observations from 1987-2000 \citep{Olah1991AA,Olah2001AA}.
The adopted $P_{\rm rot}=1.7940$ d is close to the orbital period
 $P_{\rm orb} = 1.787992$ d \citep{Halbwachs2000IAUS}. The system is in synchronous rotation,
but the photometric rotational period changes slightly due a migration of starspots over the main cycle.
The differential rotation appears to be the same as that
 on the Sun \citep{Olah1991AA,Bondar2017}, but \cite{Alekseev2005Ap} has
 proposed an anti-solar type of differential rotation.

 In modeling starspots and determining the activity level, the following general parameters
 of the star are being adopted. The spectral type of the star is usually accepted to be
 K5 \citep{Gliese1991} or K2 \citep{Oswalt1988ApJS,Olah2001AA}, however \cite{Scholz2018AA} recorded
 the spectral type as K3 IV.
Accordingly, \cite{Pettersen1989AA} observed $T_{\rm eff} =4450$ K, $R=0.77$ $R_\odot$,
 $log(L/L_\odot) = -0.68$, $M= 0.8$ $M_\odot$ \citep{Hartmann1981, Strassmeier1993},
an inclination $i = 20^\circ$ \citep{Olah1991AA}.
\cite{Olah2001AA} considered the model in which a large polar spot and some small spots on low latitudes
are present on the stellar surface and cover, in general, a significant part of the photosphere,
up to 30-50\%. \cite{Alekseev1998} derived spots areas up to 40\% assuming that
their distribution is concentrated in the equatorial zone \citep{Alekseev1997ARep}.

The long-term photometric data,  received for some decades from archives of the
 photographic measurements  \citep{Hartmann1981,Bondar1995AA}, revealed
the starspots cycle of activity of this star. The length of the cycle was estimated to be about
 60 yr and the cycle amplitude exceeds 0.6$^m$, which is an indication of a high level of photospheric activity,
 the largest among the known red dwarfs.

\section{Photometric behavior of V833 Tau in some decades and suspected activity cycles}

The specific feature of the photometric behavior of V833 Tau on
 timescale of decades is a significant change in the star brightness, up to 0.6-0.8$^ m$ in
the $V$-band relative to the unspotted maximum level \citep{Bondar2019ASPC}.
\cite{Olah2001AA} accepted $V_{\rm max}= 7.90^m$ based on observation during the 1987-2000 interval,
 and \cite{Bondar2015} found $V_{\rm max}=7.94^m$ from the long observational data over 1899-2009.
The stellar magnitude of an active star in its quiescent stage, i.e. without flares and spots,
is a very important value in order to determine parameters of starspots and to model a light curve.
This parameter can be found by using the long-term photometry or by analyzing the models of
 stellar atmospheres for a given spectral type.

The time span between the consecutive unspotted states is considered as
the length of a possible cycle of activity. To seek such stellar activity cycles,
the data rows obtained for several decades are required.
At present, the chromospheric activity of 29 stars from the MWO HK project have been monitored
 over 36 yr \citep{Olah2016AA}.
The study of photospheric activity is based on various photometric sources that
allow one to form a combined light curve over long time intervals.

The measurements by \cite{Hartmann1981} in the Harvard University's archive plate collection and
by \cite{Bondar1995AA} in the photographic archive of the Sternberg Astronomical Institute
(Moscow State University) made it possible to consider the photometric behavior of the star since 1899.
The photographic $B$-magnitudes are used to add other photometric data and to construct a common light curve.
In our studies, the combined light curve of V833 Tau based on its own and published photometric data,
the $V$-magnitudes from photometric data bases ASAS \citep{Pojmanski1997},
SuperWASP (\url{https://wasp.cerit-sc.cz/form}) and the Kamogata Wide-Field Survey
(KWS, \url{http://kws.cetus-net.org/~maehara/VSdata.py?object}) are used.

\cite{Olah2001AA} made analysis of the available photographic and photoelectric $B$ measurements from 1899 to 2000 and determined the main activity cycle to be as long as 67-69 yr.
From the more precise photoelectric light curve over 1987-2000 interval they found shorter cycles of  6.5 yr and 2.4 yr.
The authors noted that the multiplicity of cycles emphasizes a similarity between the V833 Tau and the Sun.
\cite{Bondar2015} analysed the yearly-mean $B$-magnitudes on the interval of 1899-2009
 searching for periodic changes using the statistical package AVE (\url{http://www.astrouw.edu.pl/asas/?page=acvs}).
 The results obtained allowed one to suspect a long activity cycle of 78.25 yr with
 the large amplitude of 0.6$^m$.
After the removal of this long period, the period of 18.8 yr with amplitude of 0.2-0.4$^m$ was found.
Five waves of this cycle were traced on the investigated timespan; however,
 their form, phase of minimum and amplitude change from cycle to cycle.
The residuals after accounting for the contribution of the 18.8 yr period
showed the presence of shorter cycles of 6.4 yr and 2.5 yr long \citep[similar to][]{Olah2001AA}
with amplitudes less than 0.07$^m$c. Apparently, the same cycles can be obtained from the
analysis of the $V$-light curve because the $V$-and $B$-magnitudes are related.

\begin{figure}
\includegraphics[width=1\columnwidth,right]{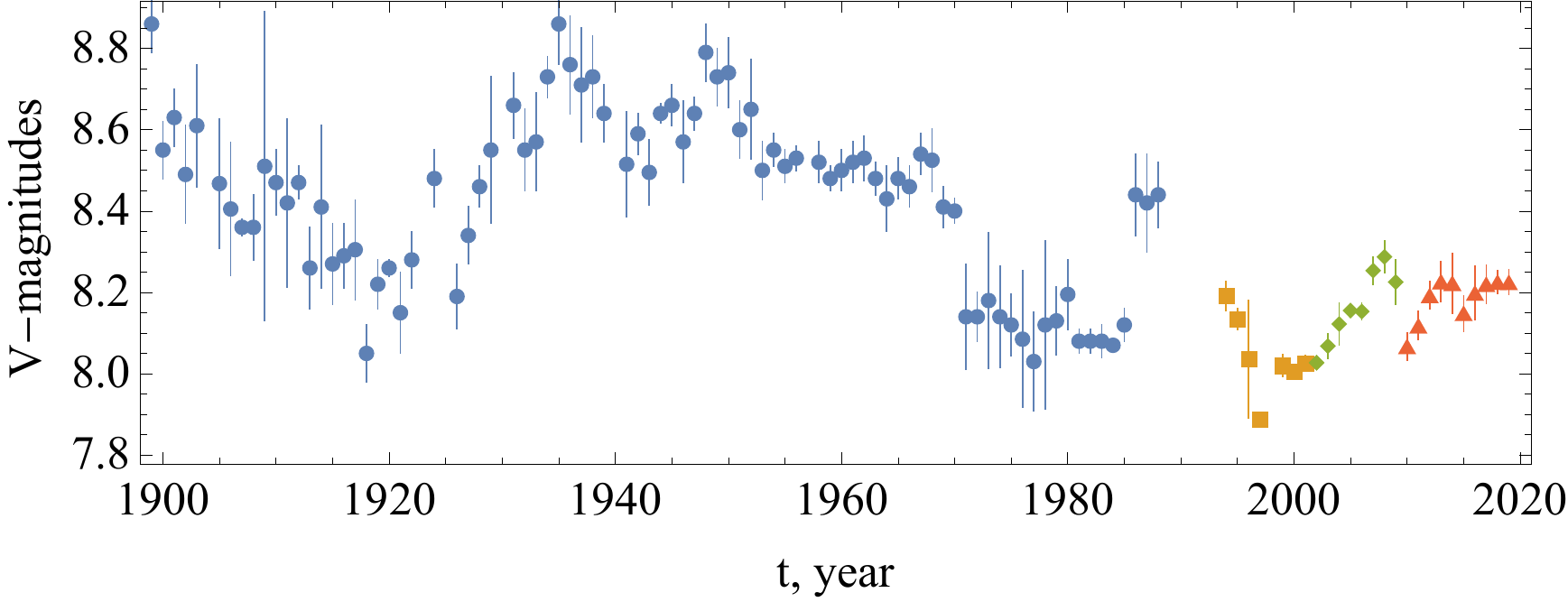}
\caption{Long-term variations of the yearly-mean V-magnitudes of the star V833 Tau
over 120~yr (1899-2019). The ordinate axis is digitized so
that the minimal brightness of a star corresponds to the maximal activity (increasing of the starspot number).
The data sets are compiled from the following sources: blue circles in the interval 1899-1988 --
\protect\cite{Hartmann1981} and \protect\cite{Bondar1995AA}, yellow squares in the interval 1994-2001 --
\protect\cite{Alekseev1998} and \protect\cite{Alekseev2001}, green rombs -- ASAS data for 2002-2009,
 red triangles -- KWS data for 2010-2919.  Error bars show standard deviations.}
    \label{fig1}
\end{figure}

In this work, we study changes in the brightness of the star over the full 1899-2019 interval.
The photographic $B$-magnitudes were transformed to $V$-magnitudes using the accepted mean value
$B-V=1.07$ \citep{Bondar2015}. The yearly-mean $V$-magnitudes were obtained as described above.
Figure~\ref{fig1} shows the light curve of the star, symbols indicate data sources,
and bars show standard error deviations. Photographic measurement errors are large,
up to $0.15^m$, and the errors of more modern photometry measurements are usually less than 0.04$^m$.

As follows from the light curve (see Fig.~\ref{fig1}), the high levels of brightness of
 V833 Tau ($V=8.00^m$) were reached during the 1918-1921 and 1976-1977 intervals, and
it was even slightly higher ($V=7.89^m$) during 1997-2000.
A remarkable decrease in the brightness began in 1928 and continued up to 1968
when the stellar magnitude was not more than 8.4$^m$, and the deep minima were observed
 around 1935 and 1950.

\section{Conservative approach for data interpretation}

Our conservative approach for the interpretation of stellar activity data
is based on applying the ideas and tools which pass all reliability tests
analysing similar solar activity records (typically, some sorts of sunspot data).
We consider the spectral properties of oscillations in the stellar spot tracer.
We suppose that stellar dynamo produces the main cycle of activity which can be identified
with a manifestation similar to the solar activity cycles.
Analysis of the pronounced peaks in the integral spectrum
is not sufficient for association with the 11-yr cycle in sunspot data.
One needs to identify how its amplitude and shape are evolving in time.
Wavelets, as a local form of the Fourier transform, seem to be
 a natural tool to get the time-frequency profile of any kinds of superposed oscillations.
Certainly, this interpretation of the 1D signal deserves a confirmation, say, in form of
a stellar butterfly diagram which should be compared with the solar ones.
However this strict demand is mainly still out of the abilities of contemporary observations.

We note that wavelet spectra for some stars do not exhibit a pronounced peak  \citep[e.g.][]{Frick1997ApJ}.
Our intention in this paper is to make one more step in this stellar-solar analogy.
An instructive example is the so-called quasi-biennial oscillations;
we bear in mind quasi-periodic oscillations in a mid-term range (i.e., ranging from 1 month to about 11 yr)
 which are emphasized in the 11-yr solar cycle at its activity maximum. However, taking into account
the whole data record and evaluating statistical significance, we have found that quasi-biennial oscillations
form a continuous spectrum of stochastic oscillations \citep{Frick2020}. This behavior reflects
the turbulent nature of solar activity.
Our wavelet analysis of a stellar activity tracer aims to recognize pronounced features
 that can be associated with the main cycle. Based on the interpretation of this similarity,
 we will do our best for now, considering what improvements can be expected given future observations.
At least, we offer a conservative first step in this study.

\subsection{Wavelet analysis}

Spectral properties of the signal which change over time can be revealed with wavelet analysis.
The continuous wavelet transform of the real signal $f(t)$ is defined as:
\begin{equation}
w_\kappa(\tau,t)={\tau}^{-\kappa}\int_{-\infty}^{\infty}f(t)\psi^*\left(\frac{t'-t}{\tau}\right)dt',
\label{wF}
\end{equation}
where $\psi(t)$ is the analysing wavelet, $\tau$  defines the scale (period of oscillation)
and $t$ defines position in time. $\kappa$ is a free parameter which is chosen
 for the sake of convenience plotting the wavelet spectrogram (balancing structures at different scale).
The signal can be restored using the inverse wavelet transform:
\begin{equation}
f(t)=\frac{1}{c_\psi}\int_{t_{\rm min}}^{t_{\rm max}}\int_{\tau_{\rm min}}^{\tau_{\rm max}} w_\kappa(\tau,t)\psi\left(\frac{t'-t}{\tau}\right)\frac{d\tau\,dt'}{{\tau}^{3-\kappa}},
\label{wFi}
\end{equation}
where $t_{\rm min}$ and $t_{\rm max}$ determine the observational period.
$\tau_{\rm min}$ and $\tau_{\rm max}$ are prescribed by the length and resolution of time series,
theoretically in an ideal case $\tau\in(0, \infty)$. In practice, these limits can depend on time.
Defining functions $\tau_{\rm min}(t)$ and $\tau_{\rm max}(t)$ allows one to restore the signal
 using a desired band of scales in the wavelet spectrogram. We use this option in the next section.
The integral wavelet spectrum (kind of a smoothed version of the Fourier spectrum) is
\begin{equation}
S(\tau)=\int_{t_{\rm min}}^{t_{\rm max}} |w_\kappa(\tau,t')|^2\frac{dt'}{{\tau}^{3-2\kappa}}.
\label{spec}
\end{equation}
The appropriate resolution in time and scale can be provided using the popular Morlet wavelet:
$
  \psi(t)=e^{-t^2} \left(e^{ 2\pi \imath t  }-e^{- \pi ^2 }\right)
$.

\subsection{Solar activity analysis}

\begin{figure}
\includegraphics[width=0.98\columnwidth,right]{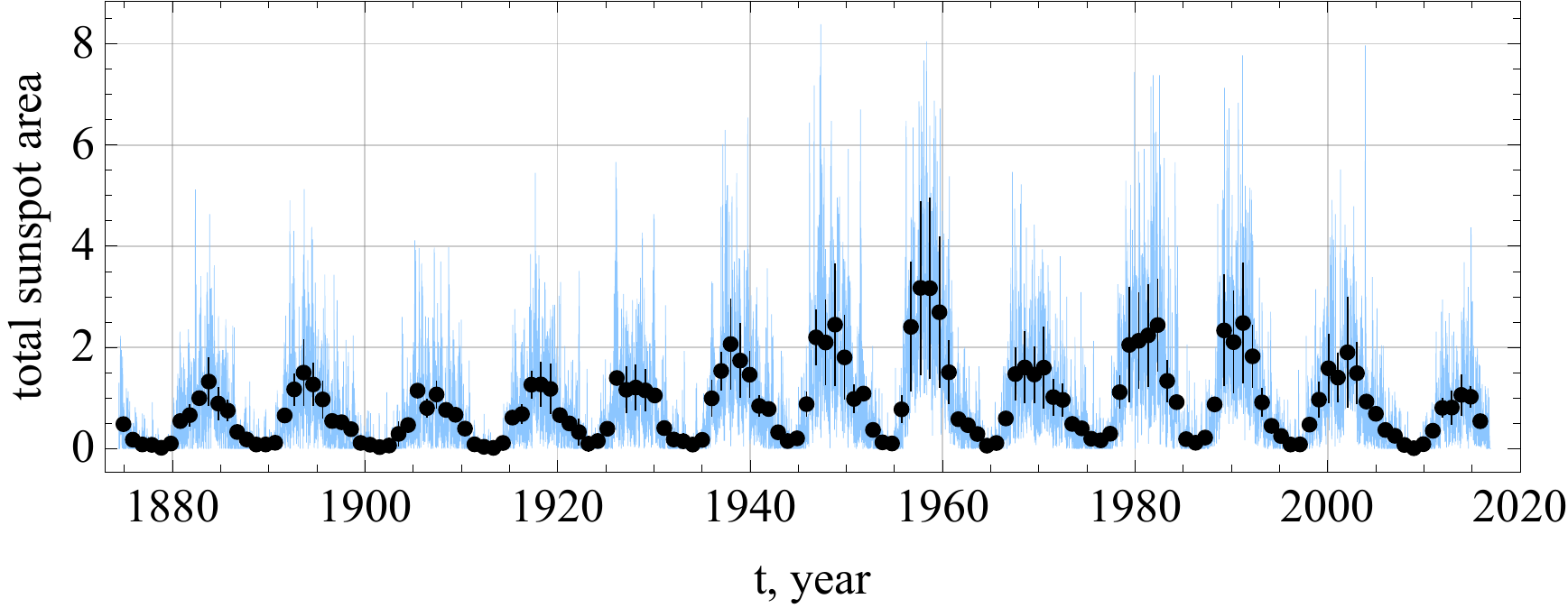}
\caption{The total sunspot area as observed by the Royal Greenwich Observatory for the period 1874-1976 and extended to 2016 by the USAF/NOAA dataset (given in units of $10^{-9}$ of solar disk). Daily variation is shown by blue curve, and yearly mean and variance is shown by black dots with bars.
} \label{fig1ss}
\end{figure}

We reproduce here the most important steps and argumentation of the following wavelet analysis
by applying them to the sunspot area record.
We choose a time interval 1876-2016 to be comparable in length
to the V833 Tau data for comparison (see in Fig.~\ref{fig1ss}).
In principle, there are more long solar activity reconstructions \citep[e.g.][]{Uetal04}
 which are very interesting for understanding of solar activity, yet they seem to be excessive
in the present context.

\begin{figure}
\includegraphics[width=1\columnwidth]{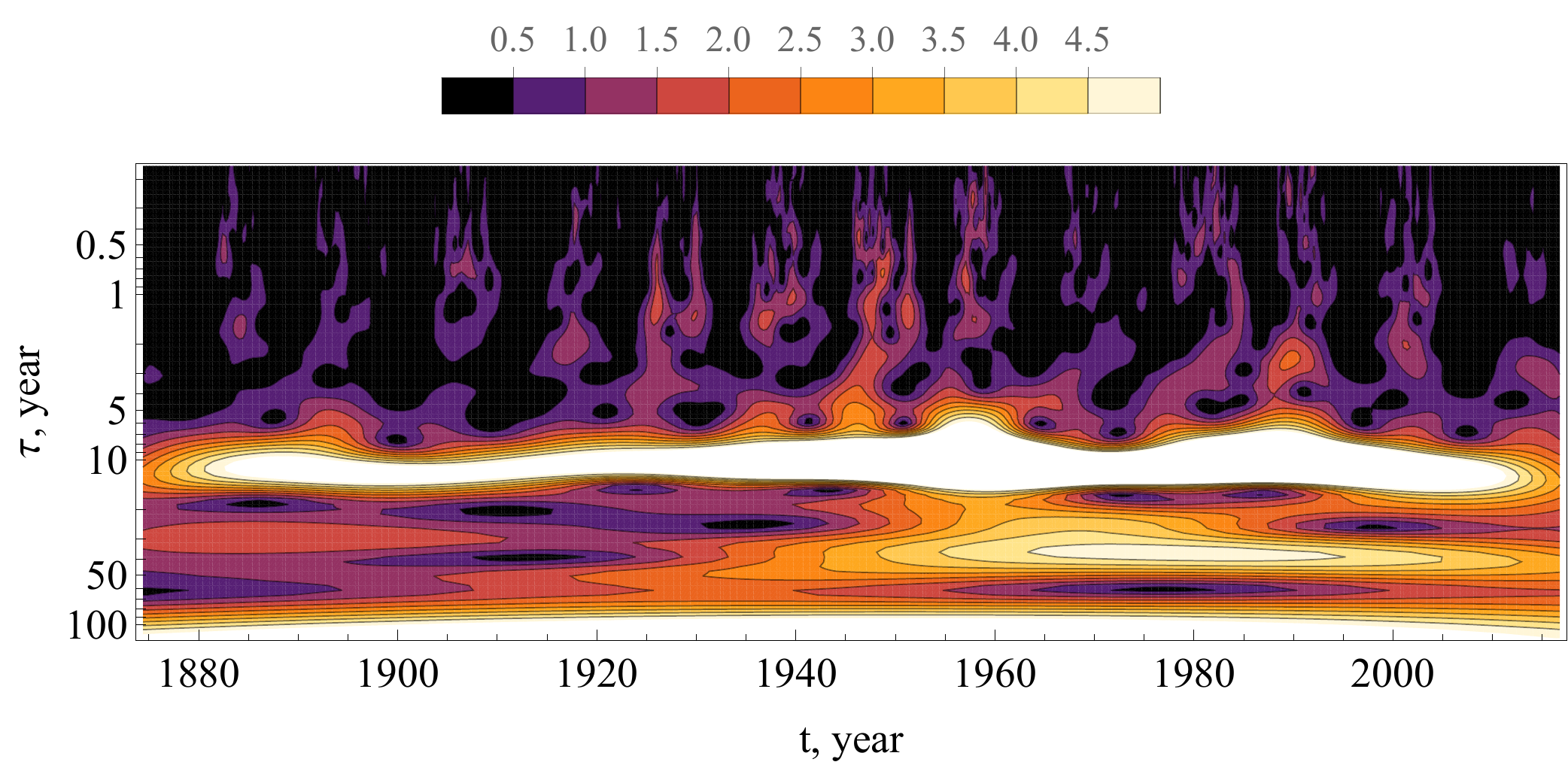}
\caption{Wavelet spectrograms for the solar data (white area denotes clipped values $|w_{1/2}|$>5)}
    \label{fig2ss}
\end{figure}

The wavelet spectrogram shown in Fig.~\ref{fig2ss} presents the intensity distribution of the wavelet
coefficients $w_{1/2}(\tau,t)$,  where the colour reflects the amplitude at the time $t$ and scale $\tau$.
The most  pronounced  strip is seen at $\tau=11$~yr and corresponds to the nominal 11-yr Schwabe cycle.
A particular wavy shape of this strip reveals amplitude and period modulations.
In fact, the amplitude of the 11-yr cycle varies significantly but we do not see it
 because the peak values are clipped and presented in white.
This allows us to show other details.
The large scale (long periodicity) structure occurs at
 $\tau\approx40$ and $t\approx1980$. The small scale (like 'waterfall') structures
occur periodically at scales $\tau<11$ and around the activity phase of maximum sunspot areas.

\begin{figure}
\includegraphics[width=0.95\columnwidth]{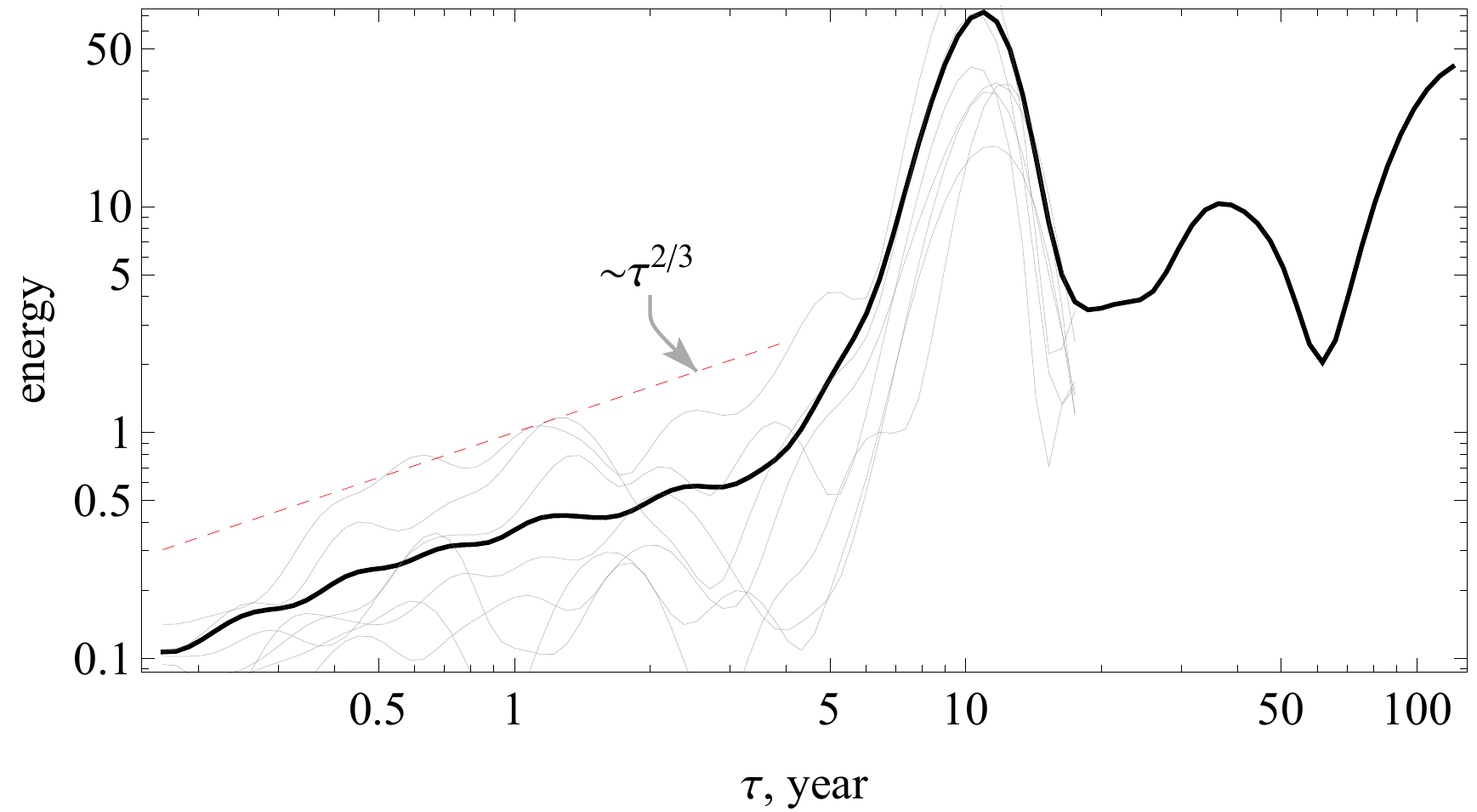}
\caption{Solar activity spectra: integral wavelet spectrum for the whole data series (thick curve) and for subsequent 22-yr intervals (thin curves).}
    \label{fig3ss}
\end{figure}

The cumulative contribution of oscillation energy at different scales can be seen in
the integral wavelet spectrum (Fig.~\ref{fig3ss}). The solar activity's 11-yr cycle is visible as
a pronounced peak nearby $\tau=11$yr. The single structure at the longer period corresponds
to a local maximum. The increase of $S(\tau)$ at $\tau=100$yr is related to the long-term variation
known as the Gleissberg cycle \citep{Gleissberg65}.
Next, we focus on the mid-term range $\tau\in[0.1,5]$ (it is in a middle between 11-yr solar cycle
and 1-month rotation period).
The overall power spectral density follows a power law with a slope of about ${2/3}$.
This means that the system exhibits stochastic behaviour under a turbulent regime.
This conclusion can be made only on the basis of a sufficiently long observation of a dozen of the
characteristic times of large-scale magnetic field evolution, in our case about 13 solar activity cycles.
 Studying individual shorter intervals, it will be difficult to confirm the interpretation
of the local peaks at any particular short periods.
One can see many of them in the individual spectra
which were obtained for six individually separated 22-yr time spans (i.e., thin curves in Fig.~\ref{fig3ss}).

\section{Wavelet analysis of the V833 Tau activity data}

\begin{figure}
\includegraphics[width=1\columnwidth]{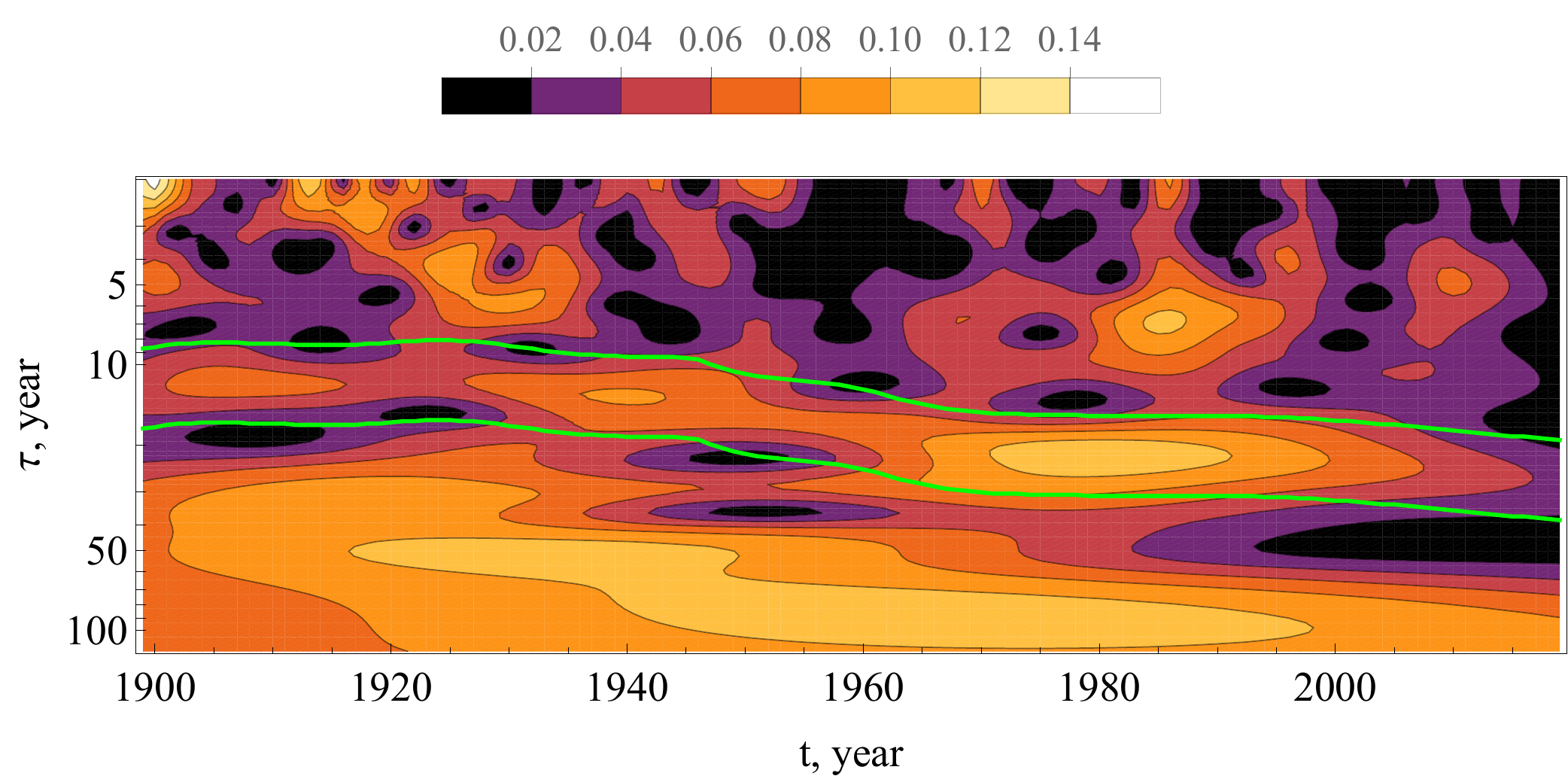}
\caption{Wavelet spectrogram for the magnetic activity of V833 Tau ($|w_{1}|$). Green lines highlight the range of scales at which the continuous oscillation is traced for further reconstruction. }
    \label{fig2}
\end{figure}

Let us consider how the proposed conservative approach would works
when applied to V833 Tau.
It looks plausible, just from the naked eye inspection of
the wavelet spectrogram in Fig.~\ref{fig2}, that V833 Tau demonstrates
a rich variety of multi-scale oscillations.
First of all, we do not see a bright strip that is in solar spectrogram (Fig.~\ref{fig2ss}).
With some care, one can distinguish a few structures between 10 and 40 yr (i.e., the interval
 where one suspects the star's dynamo may act), which are slightly linked over time.
The corresponding area is emphasised using two green lines in Fig.~\ref{fig2}.

\begin{figure}
\includegraphics[width=1\columnwidth]{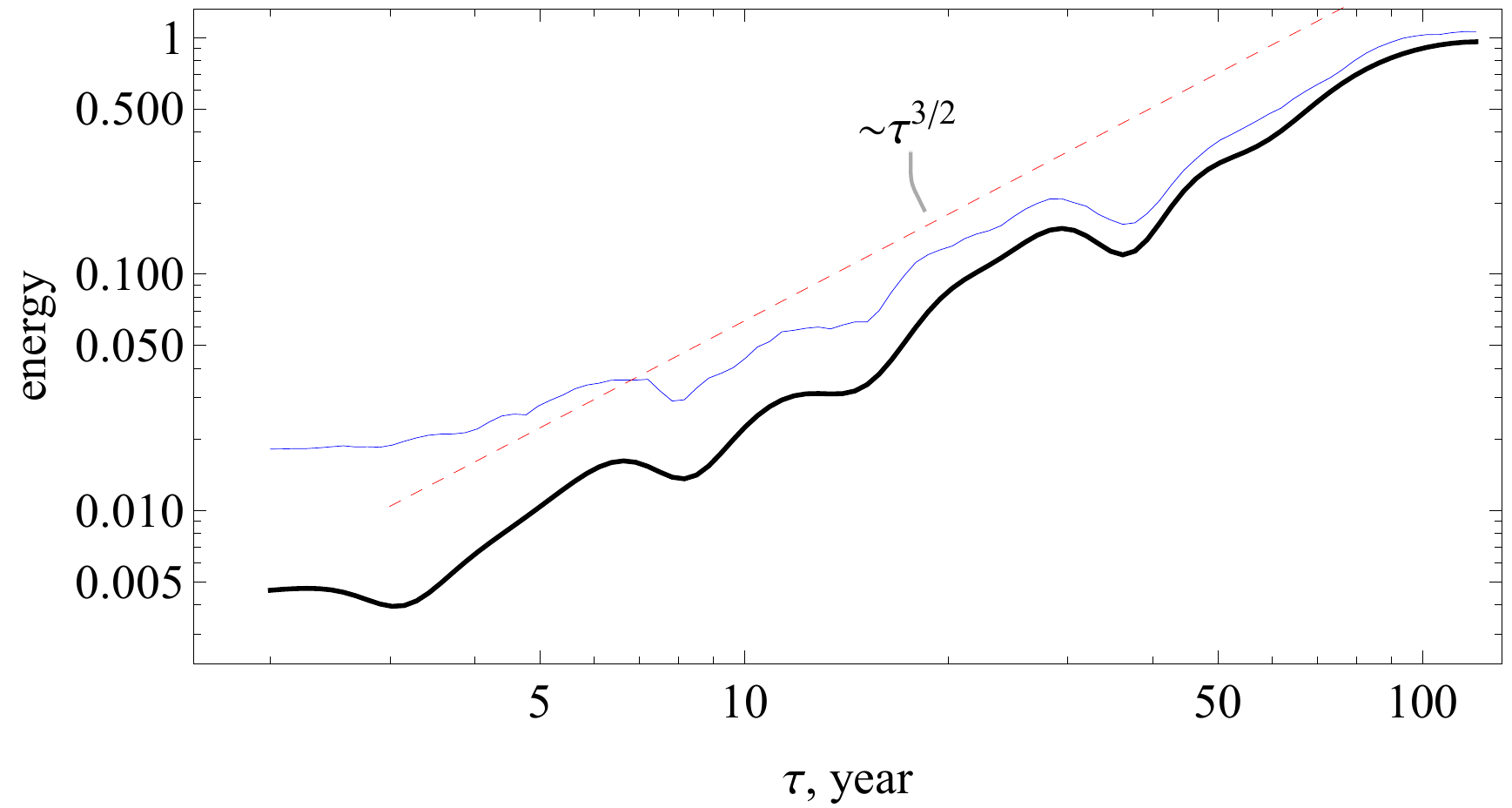}
\caption{Integral wavelet spectrum for V833 Tau (black curve). Blue thin curve shows the upper boundary of 0.9 confidential interval.}
    \label{fig3}
\end{figure}

The integral wavelet spectrum for the V833 Tau  data is shown in Fig.~\ref{fig3}.
On the one hand, we do not observe a local maximum which could be qualitatively compared
with that for the solar spectrum (Fig.~\ref{fig3ss}).
There is a bump for $\tau$ between 30 and 40 yr which is a contribution of the isolated band to the spectrogram.
We apply Monte-Carlo simulations using standard error given for each data point in order
to evaluate the upper bound of the confidential interval at the level of 0.9.
One can see that the maximum at $\tau=40$~yr is not sufficiently significant.
On the other hand, the integral wavelet spectrum of V833 Tau in Fig.~\ref{fig3} can be described by a power law
 with  $\tau^{3/2}$. The small scale part ($\tau\lesssim10$) might be affected by noise but
the rest seems statistically robust in sense that it is significant beyond observational noise.
We recall again that a power-law scaling is expected
for highly turbulent systems in the context of long observational records.
If the  statistics is still insufficient, the spectrum reveals an approximate power law relation,
and the quality of approximation should improve as more data are available.
We suppose that this has been confidently shown for solar data but it is illustrated here for V833 Tau.

\begin{figure}
\includegraphics[width=1\columnwidth]{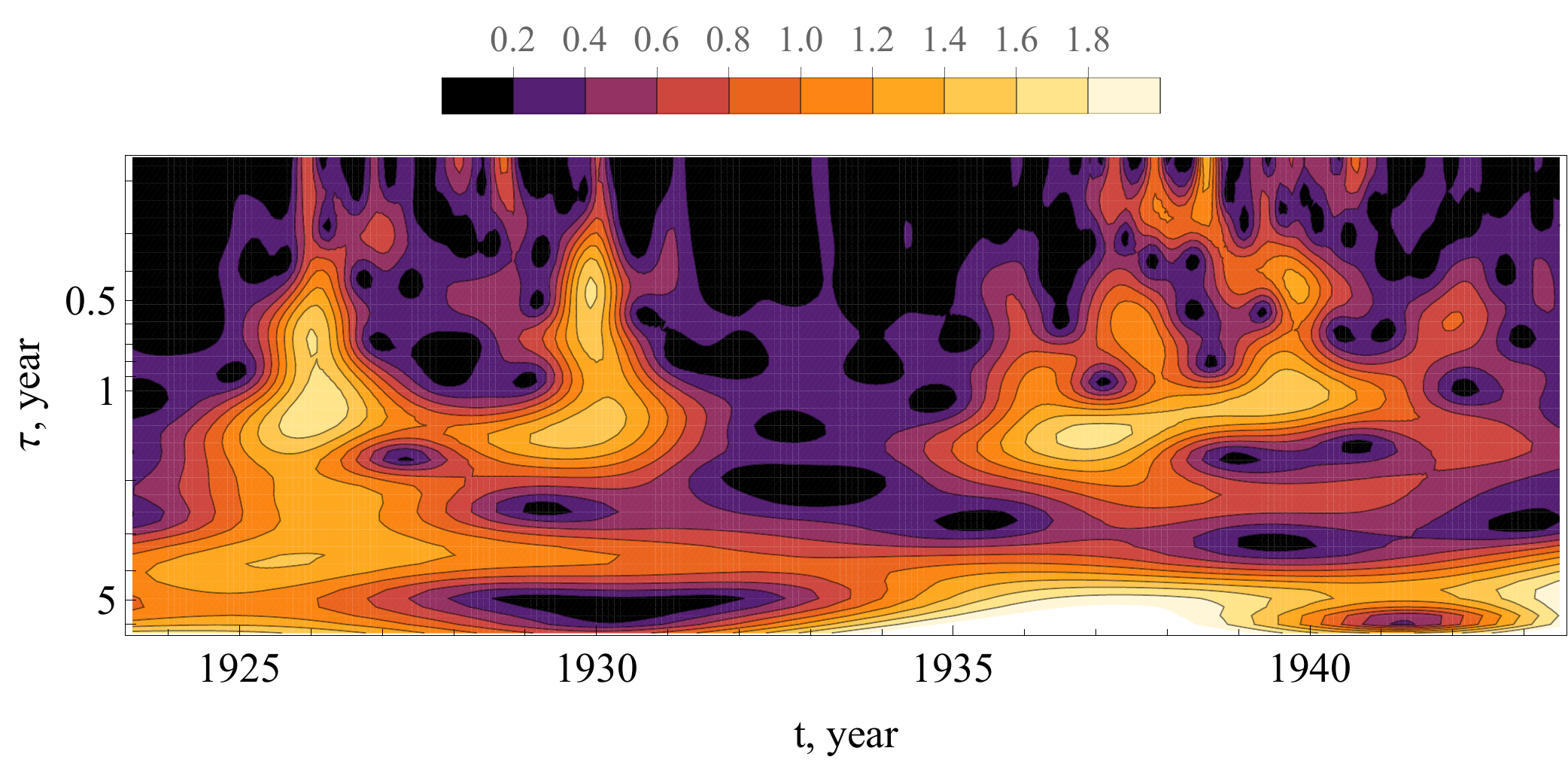}
\caption{Enlarged fragment of the wavelet spectrogram shown in Fig.~\ref{fig2ss}.}
    \label{fig5}
\end{figure}

We find that the V833 Tau spectrum follows a power-law function.
This means structural self-similarity of the V833 Tau data that is more comparable
 to the mid-term oscillations of solar activity.
Our conservative point of view necessarily leads to a conclusion that the V833 Tau activity
 cycle (expected to be similar to the solar cycle) is longer than a century, i.e.,
the timescale of V833 Tau intrinsic oscillation is an order of magnitude larger than that
 known for solar activity. This result means that observations for over about 1000 years would be needed
to demonstrate cyclic behavior for V833 Tau.
This statement can be illustrated by analysing the fragment of solar data which includes
a couple of solar cycles only.
According to this statement, we can try to see what we will observe when the solar records
 are shortened in length by 10 times. Let us examine the wavelet spectrogram for the time interval
from 1924 to 1943 (Fig.~\ref{fig5}).
We may get an impression that something important is hiding behind the set of structures
visible near $\tau=1$, or, even more important, behind the continuous strip along $\tau=5$.
 We propose that this impression would be deceptive.

\section{Discussion and conclusion}

We conclude that the V833 Tau activity data for the 120~yr between 1899-2019 obeys
the continuous spectrum of fluctuations without any significantly pronounced peaks.
The activity spectrum reveals a power law behaviour in almost the whole range of
available scales ($3<\tau<100$~yr) and this result is comparable with the behaviour of solar activity
 observed at scales $0.1 < \tau < 10$~yr. It seems natural
to suppose that V833 Tau may have a cycle of about 80-100~yr however the available data
are simply yet insufficient to isolate the proposed cycle.
 We tentatively and conservatively deduced that the physical timescale of
the dynamo engine driving the activity in V833 Tau is about 10 times longer than that for solar dynamo.

The dynamo drivers underlying solar cycle in the framework of solar dynamo include
such delicate characteristics of solar hydrodynamics as the degree of mirror asymmetry of
solar convection and the meridional circulation. Contemporary astronomers made many important attempts
 to quantify these characteristics. However, it would be an exaggeration to say that our knowledge
is sufficiently reliable in this respect, neglecting corresponding stellar properties.
In principle, stellar dynamo models admit cycles much longer than the solar ones \citep[e.g.][]{Shulyak2015MNRAS},
whereas the shorter time scale are occupied by stochastic dynamics.
The co-existing long and short stellar activity cycles can be modelled by studying
the interplay of near-surface dynamo and a dynamo operating in deeper layers
\citep{2017ApJ...845...79B}.
Analyses of the  starspot rotational modulations in 1998 main-sequence stars observed by
the Kepler satellite mission allowed one to recognize either long-living spots in activity complexes of stars
 with saturated magnetic dynamos or the spot emergence, which is modulated by large scale turbulent convection
\citep{2018MNRAS.473L..84A}.

We note that a dynamo can produce
a steady magnetic field with more or less pronounced quasi-periodic fluctuations.
Such magnetic configurations happen in spiral galaxies \citep[e.g.][]{Beck1996} or
in the Earth \citep[e.g.][]{gradstein2020}.
  It is interesting that V833 Tau is both quite young and a very active star:
its coronal radiation $L_X/L_{\rm bol}$ is close to 10$^{-3}$ and the rotation period is less than
 2 days, i.e., it rotates more than 10 times faster than the Sun, which may saturate its activity
\citep[cf.][]{Reiners2014ApJ}. For a saturated mode of activity, stellar dynamo is expected to enhance
a quasi-stationary magnetic field with chaotic variations,
yet without a pronounced cycle \citep{1996ApJ...460..848B}.
A regular cycle begins when the saturated regime of activity changes to the solar-type one,
in which the activity level depends on the rotation period \citep{2015ARep...59..726K}.
\cite{Nizamov2017AstL} showed that the activity of late spectral-type K stars
 transitions to exhibiting regular cycles when its rotation period is equal to 3 days.
It opens a possibility that V833 Tau, in its stellar evolutionary stage,
still did not reach the stage to exhibit pronounced cyclic activity.

\begin{figure}
\includegraphics[width=1\columnwidth]{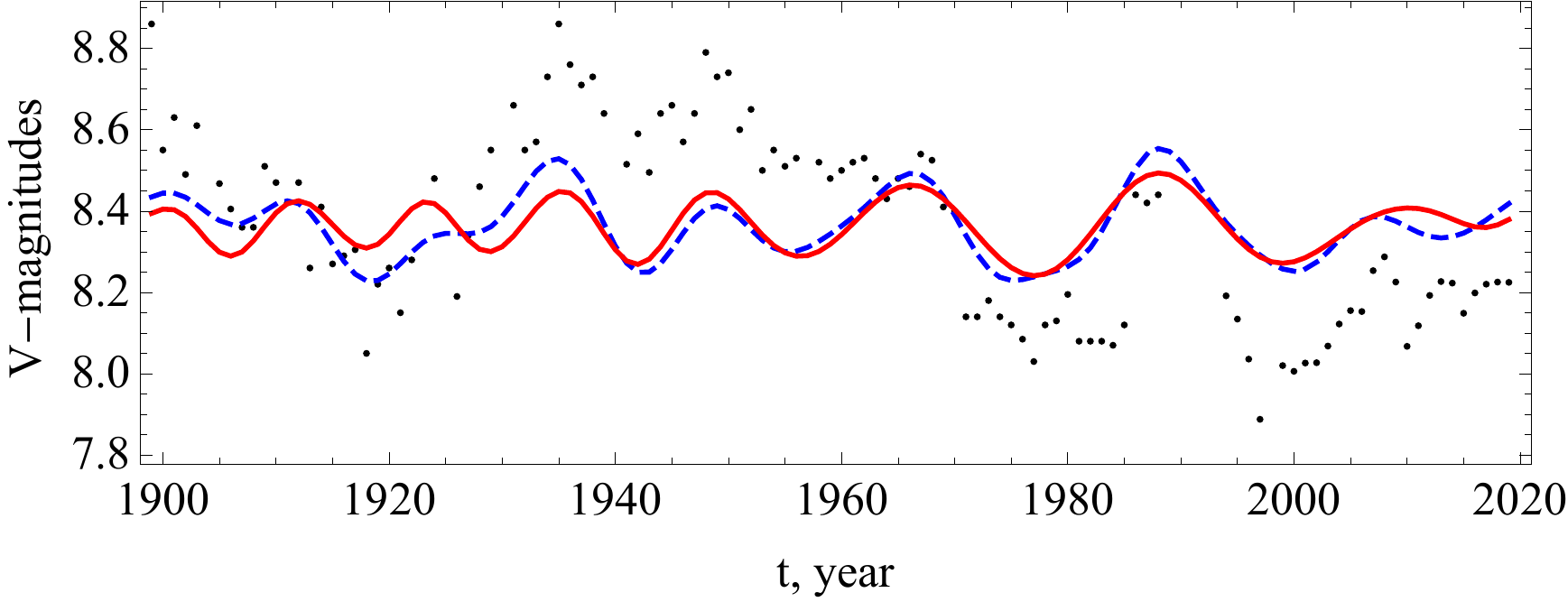}
\caption{Restored signals from the wavelet coefficients corresponding to the most pronounced spectral feature in V833 Tau data: the reconstruction along the vicinity of local maximum marked by green lines in Fig.~\ref{fig2} (red curve), and the reconstruction in the fixed scale range $10<\tau<40$~yr (blue dashed curve). Original data are shown by black points.}
    \label{fig4}
\end{figure}

An alternative point of view suggests that a time series for V833 Tau is too short
for any conclusive analysis of long-time variations ($\tau > 50$~yr) and insists
on a search for the particular oscillation in a range comparable with the Sun's 11-yr cycle.
 Following this argument, we have tried to isolate the features dominating in the range
 of timescales below 50 yrs.
We have made an attempt to restore the signal using certain wavelet coefficients corresponding
to the dynamo wave for V833 Tau.
Figure~\ref{fig4} presents two possible reconstructions,
 of which one is done using the wavelet coefficients in a fixed range of scales, i.e.,
 $10<\tau<40$~yr (blue dashed curve), and the other is done in an adaptive band as a sort of a corridor.
The latter is shown in Fig.~\ref{fig2} by green curves. At this stage, in both reconstructions,
the bright structure centered in wavelet plane at $t\approx 1980, \tau\approx 20$~yr, and match
pretty well. However, at the earlier interval, namely, for the period of
$t\approx 1920 - 1930$
the blue dashed curve has a phase break, whereas the period of oscillations traced by the red curve is
noticeably shortened (i.e., the blue curve missed one cycle against the red one; see Figure~\ref{fig4}).
Such a behavior observed in the V833 Tau record
 does not seemed to manifest itself in solar activity. In contrast, a detailed study of
  period duration and phase of the 11-yr solar oscillations
 may ultimately reveal a  synchronized clock-like process within magnetically active
 layers in the Sun \citep{1978Natur.276..676D}.

Let us summarize our main conclusions.
Wavelet analysis of the 120 yr long photometric variations for V833 Tau indicates that
the long term variations of V833 Tau are similar to mid-term solar activity.
If a stellar cycle for V833 Tau exists, then it is an order of magnitude longer than the solar one.
However, we cannot completely exclude the scenario that V833 Tau has essentially an unstable dynamo
 and its cycle length varies within the period of 10 -- 40~yr.

\section*{Acknowledgements}
We are grateful to Dr Willie Soon for his efforts to improve the paper. NB and MK acknowledge partial support by Russian Foundation for Basic Research (grant 19-02-00191a)
 and from the RFBR grant No 18-52-06002 Az-a.
DS acknowledges support from the RFBR grant No 18-02-00085.

\bibliographystyle{mnras}
\bibliography{ref,refIntro}

\begin{thebibliography}{}
\makeatletter
\relax
\def\mn@urlcharsother{\let\do\@makeother \do\$\do\&\do\#\do\^\do\_\do\%\do\~}
\def\mn@doi{\begingroup\mn@urlcharsother \@ifnextchar [ {\mn@doi@}
  {\mn@doi@[]}}
\def\mn@doi@[#1]#2{\def\@tempa{#1}\ifx\@tempa\@empty \href
  {http://dx.doi.org/#2} {doi:#2}\else \href {http://dx.doi.org/#2} {#1}\fi
  \endgroup}
\def\mn@eprint#1#2{\mn@eprint@#1:#2::\@nil}
\def\mn@eprint@arXiv#1{\href {http://arxiv.org/abs/#1} {{\tt arXiv:#1}}}
\def\mn@eprint@dblp#1{\href {http://dblp.uni-trier.de/rec/bibtex/#1.xml}
  {dblp:#1}}
\def\mn@eprint@#1:#2:#3:#4\@nil{\def\@tempa {#1}\def\@tempb {#2}\def\@tempc
  {#3}\ifx \@tempc \@empty \let \@tempc \@tempb \let \@tempb \@tempa \fi \ifx
  \@tempb \@empty \def\@tempb {arXiv}\fi \@ifundefined
  {mn@eprint@\@tempb}{\@tempb:\@tempc}{\expandafter \expandafter \csname
  mn@eprint@\@tempb\endcsname \expandafter{\@tempc}}}

\bibitem[\protect\citeauthoryear{Alekseev}{Alekseev}{2001}]{Alekseev2001}
Alekseev I.~Y.,  2001, Spotted low mass stars.
Astroprint, Odessa

\bibitem[\protect\citeauthoryear{{Alekseev}}{{Alekseev}}{2005}]{Alekseev2005Ap}
{Alekseev} I.~Y.,  2005, \mn@doi [Astrophysics] {10.1007/s10511-005-0003-x},
  \href {https://ui.adsabs.harvard.edu/abs/2005Ap.....48...20A} {48, 20}

\bibitem[\protect\citeauthoryear{{Alekseev} \& {Bondar'}}{{Alekseev} \&
  {Bondar'}}{1998}]{Alekseev1998}
{Alekseev} I.~Y.,  {Bondar'} N.~I.,  1998, Astron. Rep., \href
  {https://ui.adsabs.harvard.edu/abs/1998ARep...42..655A} {42, 655}

\bibitem[\protect\citeauthoryear{{Alekseev} \& {Gershberg}}{{Alekseev} \&
  {Gershberg}}{1997}]{Alekseev1997ARep}
{Alekseev} I.~Y.,  {Gershberg} R.~E.,  1997, Astron. Rep., \href
  {https://ui.adsabs.harvard.edu/abs/1997ARep...41..207A} {41, 207}

\bibitem[\protect\citeauthoryear{{Arkhypov}, {Khodachenko}, {Lammer},
  {G{\"u}del}, {L{\"u}ftinger}  \& {Johnstone}}{{Arkhypov}
  et~al.}{2018}]{2018MNRAS.473L..84A}
{Arkhypov} O.~V.,  {Khodachenko} M.~L.,  {Lammer} H.,  {G{\"u}del} M.,
  {L{\"u}ftinger} T.,   {Johnstone} C.~P.,  2018, \mn@doi [\mnras]
  {10.1093/mnrasl/slx170}, \href
  {https://ui.adsabs.harvard.edu/abs/2018MNRAS.473L..84A} {473, L84}

\bibitem[\protect\citeauthoryear{{Auvergne} et~al.,}{{Auvergne}
  et~al.}{2009}]{2009A&A...506..411A}
{Auvergne} M.,  et~al., 2009, \mn@doi [\aap] {10.1051/0004-6361/200810860},
  \href {https://ui.adsabs.harvard.edu/abs/2009A&A...506..411A} {506, 411}

\bibitem[\protect\citeauthoryear{{Baliunas} et~al.,}{{Baliunas}
  et~al.}{1995}]{Baliunas1995ApJ}
{Baliunas} S.~L.,  et~al., 1995, \mn@doi [\apj] {10.1086/175072}, \href
  {https://ui.adsabs.harvard.edu/abs/1995ApJ...438..269B} {438, 269}

\bibitem[\protect\citeauthoryear{{Baliunas}, {Nesme-Ribes}, {Sokoloff}  \&
  {Soon}}{{Baliunas} et~al.}{1996}]{1996ApJ...460..848B}
{Baliunas} S.~L.,  {Nesme-Ribes} E.,  {Sokoloff} D.,   {Soon} W.~H.,  1996,
  \mn@doi [\apj] {10.1086/177014}, \href
  {https://ui.adsabs.harvard.edu/abs/1996ApJ...460..848B} {460, 848}

\bibitem[\protect\citeauthoryear{{Beck}, {Brandenburg}, {Moss}, {Shukurov}  \&
  {Sokoloff}}{{Beck} et~al.}{1996}]{Beck1996}
{Beck} R.,  {Brandenburg} A.,  {Moss} D.,  {Shukurov} A.,   {Sokoloff} D.,
  1996, \mn@doi [\araa] {10.1146/annurev.astro.34.1.155}, \href
  {http://adsabs.harvard.edu/abs/1996ARA%26A..34..155B} {34, 155}

\bibitem[\protect\citeauthoryear{{Bondar'}}{{Bondar'}}{1995}]{Bondar1995AA}
{Bondar'} N.~I.,  1995, \aaps, \href
  {https://ui.adsabs.harvard.edu/abs/1995A&AS..111..259B} {111, 259}

\bibitem[\protect\citeauthoryear{{Bondar'}}{{Bondar'}}{2015}]{Bondar2015}
{Bondar'} N.~I.,  2015, \mn@doi [Astron. Rep.] {10.1134/S1063772915030014},
  \href {https://ui.adsabs.harvard.edu/abs/2015ARep...59..221B} {59, 221}

\bibitem[\protect\citeauthoryear{{Bondar'}}{{Bondar'}}{2017}]{Bondar2017}
{Bondar'} N.~I.,  2017, \mn@doi [Astron. Rep.] {10.1134/S1063772917010024},
  \href {https://ui.adsabs.harvard.edu/abs/2017ARep...61..130B} {61, 130}

\bibitem[\protect\citeauthoryear{{Bondar'}, {Gorbunov}  \&
  {Shlyapnikov}}{{Bondar'} et~al.}{2019}]{Bondar2019ASPC}
{Bondar'} N.~I.,  {Gorbunov} M.~A.,   {Shlyapnikov} A.~A.,  2019, {A Search for
  Cyclic Activity of Red Dwarfs Using Photometric Surveys}.
p.~180

\bibitem[\protect\citeauthoryear{{Bopp}, {Noah}, {Klimke}  \&
  {Africano}}{{Bopp} et~al.}{1981}]{Bopp1981ApJ}
{Bopp} B.~W.,  {Noah} P.~V.,  {Klimke} A.,   {Africano} J.,  1981, \mn@doi
  [\apj] {10.1086/159277}, \href
  {https://ui.adsabs.harvard.edu/abs/1981ApJ...249..210B} {249, 210}

\bibitem[\protect\citeauthoryear{{Borucki} et~al.,}{{Borucki}
  et~al.}{2010}]{2010Sci...327..977B}
{Borucki} W.~J.,  et~al., 2010, \mn@doi [Science] {10.1126/science.1185402},
  \href {https://ui.adsabs.harvard.edu/abs/2010Sci...327..977B} {327, 977}

\bibitem[\protect\citeauthoryear{{Brandenburg}, {Mathur}  \&
  {Metcalfe}}{{Brandenburg} et~al.}{2017}]{2017ApJ...845...79B}
{Brandenburg} A.,  {Mathur} S.,   {Metcalfe} T.~S.,  2017, \mn@doi [\apj]
  {10.3847/1538-4357/aa7cfa}, \href
  {https://ui.adsabs.harvard.edu/abs/2017ApJ...845...79B} {845, 79}

\bibitem[\protect\citeauthoryear{{Dicke}}{{Dicke}}{1978}]{1978Natur.276..676D}
{Dicke} R.~H.,  1978, \mn@doi [\nat] {10.1038/276676b0}, \href
  {https://ui.adsabs.harvard.edu/abs/1978Natur.276..676D} {276, 676}

\bibitem[\protect\citeauthoryear{{Doyle} et~al.,}{{Doyle}
  et~al.}{2018}]{2018MNRAS.475.2842D}
{Doyle} J.~G.,  et~al., 2018, \mn@doi [\mnras] {10.1093/mnras/sty032}, \href
  {https://ui.adsabs.harvard.edu/abs/2018MNRAS.475.2842D} {475, 2842}

\bibitem[\protect\citeauthoryear{{Eggen} \& {Sandage}}{{Eggen} \&
  {Sandage}}{1965}]{Eggen1965ApJ}
{Eggen} O.~J.,  {Sandage} A.~R.,  1965, \mn@doi [\apj] {10.1086/148170}, \href
  {https://ui.adsabs.harvard.edu/abs/1965ApJ...141..821E} {141, 821}

\bibitem[\protect\citeauthoryear{{Frick}, {Baliunas}, {Galyagin}, {Sokoloff}
  \& {Soon}}{{Frick} et~al.}{1997}]{Frick1997ApJ}
{Frick} P.,  {Baliunas} S.~L.,  {Galyagin} D.,  {Sokoloff} D.,   {Soon} W.,
  1997, \mn@doi [\apj] {10.1086/304206}, \href
  {https://ui.adsabs.harvard.edu/abs/1997ApJ...483..426F} {483, 426}

\bibitem[\protect\citeauthoryear{{Frick}, {Soon}, {Popova}  \&
  {Baliunas}}{{Frick} et~al.}{2004}]{Frick2004}
{Frick} P.,  {Soon} W.,  {Popova} E.,   {Baliunas} S.,  2004, \mn@doi [\na]
  {10.1016/j.newast.2004.03.005}, \href
  {https://ui.adsabs.harvard.edu/abs/2004NewA....9..599F} {9, 599}

\bibitem[\protect\citeauthoryear{{Frick}, {Sokoloff}, {Stepanov}, {Pipin}  \&
  {Usoskin}}{{Frick} et~al.}{2020}]{Frick2020}
{Frick} P.,  {Sokoloff} D.,  {Stepanov} R.,  {Pipin} V.,   {Usoskin} I.,  2020,
  \mn@doi [\mnras] {10.1093/mnras/stz3238}, \href
  {https://ui.adsabs.harvard.edu/abs/2020MNRAS.491.5572F} {491, 5572}

\bibitem[\protect\citeauthoryear{Gleissberg}{Gleissberg}{1965}]{Gleissberg65}
Gleissberg W.,  1965, J. Brit. Astr. Assoc., 75, 227

\bibitem[\protect\citeauthoryear{{Gliese} \& {Jahrei{\ss}}}{{Gliese} \&
  {Jahrei{\ss}}}{1991}]{Gliese1991}
{Gliese} W.,  {Jahrei{\ss}} H.,  1991, {Preliminary Version of the Third
  Catalogue of Nearby Stars}, On: The Astronomical Data Center CD-ROM: Selected
  Astronomical Catalogs

\bibitem[\protect\citeauthoryear{{Gradstein}, {Ogg}, Schmitz  \&
  {Ogg}}{{Gradstein} et~al.}{2020}]{gradstein2020}
{Gradstein} F.,  {Ogg} J.~G.,  Schmitz M.,   {Ogg} G.,  2020, Geologic Time
  Scale 2020.
Elsevier

\bibitem[\protect\citeauthoryear{{Griffin}, {Gunn}, {Zimmerman}  \&
  {Griffin}}{{Griffin} et~al.}{1985}]{Getal85}
{Griffin} R.~F.,  {Gunn} J.~E.,  {Zimmerman} B.~A.,   {Griffin} R.~E.~M.,
  1985, \mn@doi [\aj] {10.1086/113768}, \href
  {https://ui.adsabs.harvard.edu/abs/1985AJ.....90..609G} {90, 609}

\bibitem[\protect\citeauthoryear{{Grundahl}, {Christensen-Dalsgaard},
  {Kjeldsen}, {Frandsen}, {Arentoft}, {Kjaergaard}  \&
  {J{\o}rgensen}}{{Grundahl} et~al.}{2008}]{2008IAUS..252..465G}
{Grundahl} F.,  {Christensen-Dalsgaard} J.,  {Kjeldsen} H.,  {Frandsen} S.,
  {Arentoft} T.,  {Kjaergaard} P.,   {J{\o}rgensen} U.~G.,  2008, in {Deng} L.,
   {Chan} K.~L.,  eds,  IAU Symposium Vol. 252, The Art of Modeling Stars in
  the 21st Century. pp 465--466, \mn@doi{10.1017/S174392130802351X}

\bibitem[\protect\citeauthoryear{{Halbwachs}, {Arenou}, {Mayor}  \&
  {Udry}}{{Halbwachs} et~al.}{2000}]{Halbwachs2000IAUS}
{Halbwachs} J.-L.,  {Arenou} F.,  {Mayor} M.,   {Udry} S.,  2000, in Reipurth
  B.,  Zinnecker H.,  eds,  IAU Symposium Vol. 200, Birth and evolution of
  binary stars. p.~132

\bibitem[\protect\citeauthoryear{{Halbwachs}, {Mayor}  \& {Udry}}{{Halbwachs}
  et~al.}{2018}]{Halbwachs2018AA}
{Halbwachs} J.~L.,  {Mayor} M.,   {Udry} S.,  2018, \mn@doi [\aap]
  {10.1051/0004-6361/201833377}, \href
  {https://ui.adsabs.harvard.edu/abs/2018A&A...619A..81H} {619, A81}

\bibitem[\protect\citeauthoryear{{Hall} \& {Lockwood}}{{Hall} \&
  {Lockwood}}{1995}]{1995ApJ...438..404H}
{Hall} J.~C.,  {Lockwood} G.~W.,  1995, \mn@doi [\apj] {10.1086/175084}, \href
  {https://ui.adsabs.harvard.edu/abs/1995ApJ...438..404H} {438, 404}

\bibitem[\protect\citeauthoryear{{Hall}, {Henry}, {Lockwood}, {Skiff}  \&
  {Saar}}{{Hall} et~al.}{2009}]{2009AJ....138..312H}
{Hall} J.~C.,  {Henry} G.~W.,  {Lockwood} G.~W.,  {Skiff} B.~A.,   {Saar}
  S.~H.,  2009, \mn@doi [\aj] {10.1088/0004-6256/138/1/312}, \href
  {https://ui.adsabs.harvard.edu/abs/2009AJ....138..312H} {138, 312}

\bibitem[\protect\citeauthoryear{{Hartmann}, {Bopp}, {Dussault}, {Noah}  \&
  {Klimke}}{{Hartmann} et~al.}{1981}]{Hartmann1981}
{Hartmann} L.,  {Bopp} B.~W.,  {Dussault} M.,  {Noah} P.~V.,   {Klimke} A.,
  1981, \mn@doi [\apj] {10.1086/159326}, \href
  {https://ui.adsabs.harvard.edu/abs/1981ApJ...249..662H} {249, 662}

\bibitem[\protect\citeauthoryear{Hathaway}{Hathaway}{2015}]{hathawayLR}
Hathaway D.~H.,  2015, \mn@doi [Liv. Rev. Solar Phys.] {10.1007/lrsp-2015-4},
  12, 4

\bibitem[\protect\citeauthoryear{{Heintz}}{{Heintz}}{1981}]{Heintz1981ApJS}
{Heintz} W.~D.,  1981, \mn@doi [\apjs] {10.1086/190745}, \href
  {https://ui.adsabs.harvard.edu/abs/1981ApJS...46..247H} {46, 247}

\bibitem[\protect\citeauthoryear{{Karak}, {Tomar}  \& {Vashishth}}{{Karak}
  et~al.}{2020}]{2020MNRAS.491.3155K}
{Karak} B.~B.,  {Tomar} A.,   {Vashishth} V.,  2020, \mn@doi [\mnras]
  {10.1093/mnras/stz3220}, \href
  {https://ui.adsabs.harvard.edu/abs/2020MNRAS.491.3155K} {491, 3155}

\bibitem[\protect\citeauthoryear{{Karoff}, {Metcalfe}, {Montet}, {Jannsen},
  {Santos}, {Nielsen}  \& {Chaplin}}{{Karoff}
  et~al.}{2019}]{2019MNRAS.485.5096K}
{Karoff} C.,  {Metcalfe} T.~S.,  {Montet} B.~T.,  {Jannsen} N.~E.,  {Santos}
  A.~R.~G.,  {Nielsen} M.~B.,   {Chaplin} W.~J.,  2019, \mn@doi [\mnras]
  {10.1093/mnras/stz782}, \href
  {https://ui.adsabs.harvard.edu/abs/2019MNRAS.485.5096K} {485, 5096}

\bibitem[\protect\citeauthoryear{{Katsova}, {Bondar}  \& {Livshits}}{{Katsova}
  et~al.}{2015}]{2015ARep...59..726K}
{Katsova} M.~M.,  {Bondar} N.~I.,   {Livshits} M.~A.,  2015, \mn@doi [Astron.
  Rep.] {10.1134/S1063772915070045}, \href
  {https://ui.adsabs.harvard.edu/abs/2015ARep...59..726K} {59, 726}

\bibitem[\protect\citeauthoryear{{Kiefer}, {Schad}, {Davies}  \&
  {Roth}}{{Kiefer} et~al.}{2017}]{2017A&A...598A..77K}
{Kiefer} R.,  {Schad} A.,  {Davies} G.,   {Roth} M.,  2017, \mn@doi [\aap]
  {10.1051/0004-6361/201628469}, \href
  {https://ui.adsabs.harvard.edu/abs/2017A&A...598A..77K} {598, A77}

\bibitem[\protect\citeauthoryear{{Kiefer}, {Broomhall}  \& {Ball}}{{Kiefer}
  et~al.}{2019}]{2019FrASS...6...52K}
{Kiefer} R.,  {Broomhall} A.-M.,   {Ball} W.~H.,  2019, \mn@doi [Front. Astron.
  Space Sci.] {10.3389/fspas.2019.00052}, \href
  {https://ui.adsabs.harvard.edu/abs/2019FrASS...6...52K} {6, 52}

\bibitem[\protect\citeauthoryear{{Lehmann}, {Jardine}, {Mackay}  \&
  {Vidotto}}{{Lehmann} et~al.}{2018}]{2018MNRAS.478.4390L}
{Lehmann} L.~T.,  {Jardine} M.~M.,  {Mackay} D.~H.,   {Vidotto} A.~A.,  2018,
  \mn@doi [\mnras] {10.1093/mnras/sty1230}, \href
  {https://ui.adsabs.harvard.edu/abs/2018MNRAS.478.4390L} {478, 4390}

\bibitem[\protect\citeauthoryear{{Lockwood}, {Skiff}  \& {Radick}}{{Lockwood}
  et~al.}{1997}]{1997ApJ...485..789L}
{Lockwood} G.~W.,  {Skiff} B.~A.,   {Radick} R.~R.,  1997, \mn@doi [\apj]
  {10.1086/304453}, \href
  {https://ui.adsabs.harvard.edu/abs/1997ApJ...485..789L} {485, 789}

\bibitem[\protect\citeauthoryear{{Lucas} \& {Roche}}{{Lucas} \&
  {Roche}}{2002}]{LucasRoche2002}
{Lucas} P.~W.,  {Roche} P.~F.,  2002, \mn@doi [\mnras]
  {10.1046/j.1365-8711.2002.05786.x}, \href
  {https://ui.adsabs.harvard.edu/abs/2002MNRAS.336..637L} {336, 637}

\bibitem[\protect\citeauthoryear{{Montet}, {Tovar}  \&
  {Foreman-Mackey}}{{Montet} et~al.}{2017}]{2017ApJ...851..116M}
{Montet} B.~T.,  {Tovar} G.,   {Foreman-Mackey} D.,  2017, \mn@doi [\apj]
  {10.3847/1538-4357/aa9e00}, \href
  {https://ui.adsabs.harvard.edu/abs/2017ApJ...851..116M} {851, 116}

\bibitem[\protect\citeauthoryear{{Moss} \& {Sokoloff}}{{Moss} \&
  {Sokoloff}}{2007}]{2007MNRAS.377.1597M}
{Moss} D.,  {Sokoloff} D.,  2007, \mn@doi [\mnras]
  {10.1111/j.1365-2966.2007.11721.x}, \href
  {https://ui.adsabs.harvard.edu/abs/2007MNRAS.377.1597M} {377, 1597}

\bibitem[\protect\citeauthoryear{{Moss}, {Sokoloff}  \& {Lanza}}{{Moss}
  et~al.}{2011}]{2011A&A...531A..43M}
{Moss} D.,  {Sokoloff} D.,   {Lanza} A.~F.,  2011, \mn@doi [\aap]
  {10.1051/0004-6361/201015949}, \href
  {https://ui.adsabs.harvard.edu/abs/2011A&A...531A..43M} {531, A43}

\bibitem[\protect\citeauthoryear{{Nizamov}, {Katsova}  \& {Livshits}}{{Nizamov}
  et~al.}{2017}]{Nizamov2017AstL}
{Nizamov} B.~A.,  {Katsova} M.~M.,   {Livshits} M.~A.,  2017, \mn@doi
  [Astronomy Letters] {10.1134/S1063773717020049}, \href
  {https://ui.adsabs.harvard.edu/abs/2017AstL...43..202N} {43, 202}

\bibitem[\protect\citeauthoryear{{Olah} \& {Pettersen}}{{Olah} \&
  {Pettersen}}{1991}]{Olah1991AA}
{Olah} K.,  {Pettersen} B.~R.,  1991, \aap, \href
  {https://ui.adsabs.harvard.edu/abs/1991A&A...242..443O} {242, 443}

\bibitem[\protect\citeauthoryear{{Ol{\'a}h}, {Strassmeier}, {Kov{\'a}ri}  \&
  {Guinan}}{{Ol{\'a}h} et~al.}{2001}]{Olah2001AA}
{Ol{\'a}h} K.,  {Strassmeier} K.~G.,  {Kov{\'a}ri} Z.,   {Guinan} E.~F.,  2001,
  \mn@doi [\aap] {10.1051/0004-6361:20010362}, \href
  {https://ui.adsabs.harvard.edu/abs/2001A&A...372..119O} {372, 119}

\bibitem[\protect\citeauthoryear{{Ol{\'a}h}, {K{\H{o}}v{\'a}ri}, {Petrovay},
  {Soon}, {Baliunas}, {Koll{\'a}th}  \& {Vida}}{{Ol{\'a}h}
  et~al.}{2016}]{Olah2016AA}
{Ol{\'a}h} K.,  {K{\H{o}}v{\'a}ri} Z.,  {Petrovay} K.,  {Soon} W.,  {Baliunas}
  S.,  {Koll{\'a}th} Z.,   {Vida} K.,  2016, \mn@doi [\aap]
  {10.1051/0004-6361/201628479}, \href
  {https://ui.adsabs.harvard.edu/abs/2016A&A...590A.133O} {590, A133}

\bibitem[\protect\citeauthoryear{{Oswalt}, {Hintzen}  \& {Luyten}}{{Oswalt}
  et~al.}{1988}]{Oswalt1988ApJS}
{Oswalt} T.~D.,  {Hintzen} P.~M.,   {Luyten} W.~J.,  1988, \mn@doi [\apjs]
  {10.1086/191263}, \href
  {https://ui.adsabs.harvard.edu/abs/1988ApJS...66..391O} {66, 391}

\bibitem[\protect\citeauthoryear{{Parker}}{{Parker}}{1955}]{Parker55}
{Parker} E.~N.,  1955, \mn@doi [\apj] {10.1086/146087}, \href
  {https://ui.adsabs.harvard.edu/abs/1955ApJ...122..293P} {122, 293}

\bibitem[\protect\citeauthoryear{{Pettersen}}{{Pettersen}}{1989}]{Pettersen1989AA}
{Pettersen} B.~R.,  1989, \aap, \href
  {https://ui.adsabs.harvard.edu/abs/1989A&A...209..279P} {209, 279}

\bibitem[\protect\citeauthoryear{{Pojmanski}}{{Pojmanski}}{1997}]{Pojmanski1997}
{Pojmanski} G.,  1997, \actaa, \href
  {https://ui.adsabs.harvard.edu/abs/1997AcA....47..467P} {47, 467}

\bibitem[\protect\citeauthoryear{{Radick}, {Lockwood}, {Skiff}  \&
  {Baliunas}}{{Radick} et~al.}{1998}]{1998ApJS..118..239R}
{Radick} R.~R.,  {Lockwood} G.~W.,  {Skiff} B.~A.,   {Baliunas} S.~L.,  1998,
  \mn@doi [\apjs] {10.1086/313135}, \href
  {https://ui.adsabs.harvard.edu/abs/1998ApJS..118..239R} {118, 239}

\bibitem[\protect\citeauthoryear{{Radick}, {Lockwood}, {Henry}, {Hall}  \&
  {Pevtsov}}{{Radick} et~al.}{2018}]{2018ApJ...855...75R}
{Radick} R.~R.,  {Lockwood} G.~W.,  {Henry} G.~W.,  {Hall} J.~C.,   {Pevtsov}
  A.~A.,  2018, \mn@doi [\apj] {10.3847/1538-4357/aaaae3}, \href
  {https://ui.adsabs.harvard.edu/abs/2018ApJ...855...75R} {855, 75}

\bibitem[\protect\citeauthoryear{{Reiners}, {Sch{\"u}ssler}  \&
  {Passegger}}{{Reiners} et~al.}{2014}]{Reiners2014ApJ}
{Reiners} A.,  {Sch{\"u}ssler} M.,   {Passegger} V.~M.,  2014, \mn@doi [\apj]
  {10.1088/0004-637X/794/2/144}, \href
  {https://ui.adsabs.harvard.edu/abs/2014ApJ...794..144R} {794, 144}

\bibitem[\protect\citeauthoryear{{Salabert} et~al.,}{{Salabert}
  et~al.}{2016}]{2016A&A...596A..31S}
{Salabert} D.,  et~al., 2016, \mn@doi [\aap] {10.1051/0004-6361/201628583},
  \href {https://ui.adsabs.harvard.edu/abs/2016A&A...596A..31S} {596, A31}

\bibitem[\protect\citeauthoryear{{Scholz}, {Meusinger}  \&
  {Jahrei{\ss}}}{{Scholz} et~al.}{2018}]{Scholz2018AA}
{Scholz} R.~D.,  {Meusinger} H.,   {Jahrei{\ss}} H.,  2018, \mn@doi [\aap]
  {10.1051/0004-6361/201731753}, \href
  {https://ui.adsabs.harvard.edu/abs/2018A&A...613A..26S} {613, A26}

\bibitem[\protect\citeauthoryear{{Schrijver} \& {Zwaan}}{{Schrijver} \&
  {Zwaan}}{2008}]{2008ssma.book.....S}
{Schrijver} C.~J.,  {Zwaan} C.,  2008, {Solar and Stellar Magnetic Activity}.
Cambridge University Press

\bibitem[\protect\citeauthoryear{{Shulyak}, {Sokoloff}, {Kitchatinov}  \&
  {Moss}}{{Shulyak} et~al.}{2015}]{Shulyak2015MNRAS}
{Shulyak} D.,  {Sokoloff} D.,  {Kitchatinov} L.,   {Moss} D.,  2015, \mn@doi
  [\mnras] {10.1093/mnras/stv585}, \href
  {https://ui.adsabs.harvard.edu/abs/2015MNRAS.449.3471S} {449, 3471}

\bibitem[\protect\citeauthoryear{{Strassmeier}, {Hall}, {Fekel}  \&
  {Scheck}}{{Strassmeier} et~al.}{1993}]{Strassmeier1993}
{Strassmeier} K.~G.,  {Hall} D.~S.,  {Fekel} F.~C.,   {Scheck} M.,  1993,
  \aaps, \href {https://ui.adsabs.harvard.edu/abs/1993A&AS..100..173S} {100,
  173}

\bibitem[\protect\citeauthoryear{{Testa}, {Saar}  \& {Drake}}{{Testa}
  et~al.}{2015}]{Testa15}
{Testa} P.,  {Saar} S.~H.,   {Drake} J.~J.,  2015, \mn@doi [Phil. Trans. R.
  Soc. A] {10.1098/rsta.2014.0259}, \href
  {https://ui.adsabs.harvard.edu/abs/2015RSPTA.37340259T} {373, 20140259}

\bibitem[\protect\citeauthoryear{{Usoskin}, {Mursula}, {Solanki},
  {Sch{\"u}ssler}  \& {Alanko}}{{Usoskin} et~al.}{2004}]{Uetal04}
{Usoskin} I.~G.,  {Mursula} K.,  {Solanki} S.,  {Sch{\"u}ssler} M.,   {Alanko}
  K.,  2004, \mn@doi [\aap] {10.1051/0004-6361:20031533}, \href
  {https://ui.adsabs.harvard.edu/abs/2004A&A...413..745U} {413, 745}

\bibitem[\protect\citeauthoryear{{Vida} et~al.,}{{Vida}
  et~al.}{2016}]{2016A&A...590A..11V}
{Vida} K.,  et~al., 2016, \mn@doi [\aap] {10.1051/0004-6361/201527925}, \href
  {https://ui.adsabs.harvard.edu/abs/2016A&A...590A..11V} {590, A11}

\bibitem[\protect\citeauthoryear{{Vidotto} et~al.,}{{Vidotto}
  et~al.}{2014}]{2014MNRAS.441.2361V}
{Vidotto} A.~A.,  et~al., 2014, \mn@doi [\mnras] {10.1093/mnras/stu728}, \href
  {https://ui.adsabs.harvard.edu/abs/2014MNRAS.441.2361V} {441, 2361}

\bibitem[\protect\citeauthoryear{{Viviani}, {Warnecke}, {K{\"a}pyl{\"a}},
  {K{\"a}pyl{\"a}}, {Olspert}, {Cole-Kodikara}, {Lehtinen}  \&
  {Brandenburg}}{{Viviani} et~al.}{2018}]{2018A&A...616A.160V}
{Viviani} M.,  {Warnecke} J.,  {K{\"a}pyl{\"a}} M.~J.,  {K{\"a}pyl{\"a}} P.~J.,
   {Olspert} N.,  {Cole-Kodikara} E.~M.,  {Lehtinen} J.~J.,   {Brandenburg} A.,
   2018, \mn@doi [\aap] {10.1051/0004-6361/201732191}, \href
  {https://ui.adsabs.harvard.edu/abs/2018A&A...616A.160V} {616, A160}

\bibitem[\protect\citeauthoryear{{Walker} et~al.,}{{Walker}
  et~al.}{2003}]{2003PASP..115.1023W}
{Walker} G.,  et~al., 2003, \mn@doi [\pasp] {10.1086/377358}, \href
  {https://ui.adsabs.harvard.edu/abs/2003PASP..115.1023W} {115, 1023}

\makeatother
\end{thebibliography}

\bsp    
\label{lastpage}
\end{document}